\documentclass[pra,twocolumn,aps,superscriptaddress,longbibliography]{revtex4-1}

\usepackage[normalem]{ulem}
\usepackage{amsmath}
\usepackage{amssymb}
\usepackage{lineno}
\usepackage{graphicx}
\usepackage[usenames,dvipsnames]{color}
\usepackage{braket}
\usepackage{bm}
\usepackage{mathtools}
\usepackage{times}
\usepackage{hyperref}
\usepackage{pdfpages}
\usepackage{float}
\usepackage{lipsum}
\usepackage[dvipsnames]{xcolor}
\hypersetup{
  colorlinks=true,
  citecolor=blue,
  linkcolor=blue,
  urlcolor=blue}
\usepackage{tikz}
\usetikzlibrary{decorations.markings}
\usetikzlibrary{decorations.pathmorphing,snakes}
\usetikzlibrary{arrows.meta}
\tikzset{middlearrow/.style={
        decoration={markings,
            mark= at position 0.5 with {\arrow{#1}} ,
        },
        postaction={decorate}
    }
}

\makeatletter
 \AtBeginDocument{\let\LS@rot\@undefined}
\makeatother

\begin{document}
\title{Quench dynamics across the MI-SF quantum phase transition 
       with cluster mean field theory }

\author{Deepak Gaur}
\affiliation{Physical Research Laboratory,
             Ahmedabad - 380009, Gujarat,
             India}
\affiliation{Indian Institute of Technology Gandhinagar,
             Palaj, Gandhinagar - 382355, Gujarat,
             India}

\author{Hrushikesh Sable}
\affiliation{Physical Research Laboratory,
             Ahmedabad - 380009, Gujarat,
             India}
\affiliation{Department of Physics, Virginia Tech,
             Blacksburg, Virginia 24061, USA}

\author{D. Angom}
\affiliation{Department of Physics, Manipur University,
             Canchipur - 795003, Manipur,
             India}

\begin{abstract}
In this work, we study the quench dynamics of quantum phases of ultracold
neutral bosons trapped in optical lattices. We investigate the validity of the
Kibble-Zurek (KZ) scaling laws with the single-site Gutzwiller mean-field (SGMF)
and cluster Gutzwiller mean-field (CGMF) theory. With CGMF, we note the 
evolution of the dynamical wavefunction in the ``impulse" regime of the
Kibble-Zurek mechanism. We obtain the power law scalings for the crossover time
and defect density with the quench rate predicted by KZ scaling laws.
The critical exponents obtained from dynamics are close to their equilibrium 
values. Furthermore, it is observed that the obtained dynamical critical
exponent $z$ improves towards the equilibrium value with increasing cluster sizes in CGMF.

\end{abstract}

\maketitle


\section{Introduction}
The non-equilibrium phenomena involving a quench across a continuous phase
transition follow universal scalings relations given in terms of the
equilibrium critical exponents, as explained by the Kibble-Zurek Mechanism
(KZM) \cite{zurek_85, zurek_05, dziarmaga_05, delCampo_14}.
The KZM is based on the idea of critical slowing down near the continuous phase
transition. As a result, the system is no longer able to catch up the
change in the parameters, however slow the change in parameters be, and the
adiabaticity breaks down. The spontaneous symmetry breaking at the phase
transition would then lead to topological defects in the system due to
independent local choices of symmetry breaking in different parts of the
system. KZM assumes the temporal evolution of the state as adiabatic far from
the critical point and assumes it as an impulse when the system parameters
are sufficiently close to the critical point of phase transition. After the
critical point is passed and the system parameters are sufficiently away, the
adiabaticity is finally restored. With this, the KZM predicts the density of
the topological defects to follow a power law scaling with the quench rate,
and the scaling exponents are given in terms of the equilibrium critical
exponents of the phase transition.
In Condensed matter systems, the KZM started with the works of Zurek, and the
prediction of the scaling laws for the correlation length and defect density
was afterward experimentally verified in superfluid helium
\cite{zurek_85, zurek_96, ruutu_96, bauerle_96}. Later on, the KZM was applied
to classical phase transitions in a variety of condensed matter systems and
eventually to quantum phase transitions (QPTs) \cite{zurek_05, dziarmaga_12,
delCampo_14}. 

Ultracold neutral bosonic atoms in optical lattices are excellent systems for
studying quantum phases and the QPTs and quantum phases in the laboratory.
This engineered system is used as a macroscopic quantum simulator for a variety
of condensed matter systems \cite{lewenstein_07, gross_17}.
We investigate the quench dynamics of the quantum phase supported by the
system of ultracold neutral bosons in optical lattices described by the
Bose-Hubbard model (BHM). The ground state of the BHM Hamiltonian consists of
the incompressible Mott-insulator (MI) and compressible superfluid (SF) states
as the quantum phases \cite{fisher_89, jaksch_98}. The BHM shows a QPT between
the MI and SF phases at the critical value of the model parameters 
\cite{greiner_02_1, greiner_02_2}.
We have studied the KZM scaling laws for MI-SF QPT by evolving the MI state
under a quantum quench, taking the system across the phase transition into
the SF regime. Various experimental works exist on the quench dynamics
across the MI-SF QPT and KZ scaling \cite{chen_11, braun_15}.
On the theoretical side, there are studies investigating the KZM for QPTs in BHM
and its extensions \cite{weiss_18, shimizu_misf_18, shimizu_dwss_18,
shimizu_dwsf_18,zhou_20, hrushi_21, sable_21, pardeep_23}. However, the 
dynamically obtained exponents differ from the equilibrium critical exponents.
In this work, we explore the KZM scaling laws with the SGMF method and compare
it with the CGMF study to see the effect due to intra-cluster dynamics, which is
captured exactly compared to the single-site mean-field description. The CGMF
also allows better relaxation dynamics due to a large Fock-state space.
We have chosen to study the QPT at fixed chemical potential $\mu = 0.3U$, 
which corresponds to a first-order phase transition, and also at the tip of the
Mott-1 lobe at $\mu = 0.4U$, which corresponds to a continuous phase transition 
and belongs to the 3D XY model universality class. The quench is performed by
increasing the hopping strength. For MI-SF QPT, the equilibrium critical
exponents have the values of the 3D XY model at the tip of the Mott lob
$\nu = 2/3$ and $z = 1,$ and away from the tip, the exponents have mean-field
values $\nu=1/2$ and $z =2$ \cite{fisher_89}.
In our study, we observe the power-law
scalings as predicted by KZM and find that the dynamically obtained critical 
exponents deviate from their equilibrium values. However, the mismatch for the
critical exponent associated with the divergence of relaxation time reduces
with the CGMF study. The CGMF study captures the evolution in the quenched state
in the ``impulse" regime of KZM against the assumption of frozen dynamics.
This dynamics is absent in the SGMF studies. The frozen state dynamics in the
impulse regime have been explored in various works 
\cite{biroli_10, jeong_19, krishanu_21}. The many-body localization,
thermalization, and entanglement dynamics following the quench are various other
research pursuits explored using the quantum quenches
\cite{polkonikov_11, abanin_19}.

We have organized the remainder of this article as follows.
In Sec.{\ref{sec_theory}}, we describe the BHM model, which describes the
system. Here, we also give a brief discussion of the numerical mean-field
methods used by us and a brief discussion of KZM. This is followed by the
results from the quench dynamics study, which is discussed in
Sec.{\ref{sec_results}}. 
We first discuss the quench dynamics across the QPT below the tip of the Mott
lobe. Here, we discuss the findings from SGMF and CGMF studies and compare them.
At the end, we discuss the quench across the QPT at the tip of the Mott lobe.
Finally, we summarize and conclude the findings in Sec.{\ref{sec_summary}}

\section{Theory} \label{sec_theory}

Consider a system of bosonic atoms loaded in a 2D square optical lattice. 
The system is well described by the Bose-Hubbard model and the Hamiltonian
of the system is \cite{jaksch_98}
\begin{eqnarray}
   \hat{H} = \sum_{p, q}\bigg [ &&-\Big( J
             \hat{b}^{\dagger}_{p+1, q}\hat{b}_{p, q} 
             + J \hat{b}^{\dagger}_{p, q+1}\hat{b}_{p, q} + {\rm H.c.}\Big)
             \nonumber\\ 
             &&+ \frac{U}{2} \hat{n}_{p, q} (\hat{n}_{p, q}-1) - 
             \mu\hat{n}_{p, q}\bigg]
  \label{ham}  
\end{eqnarray}
where $p$ ($q$) is the lattice site index along the $x$ ($y$) direction, 
$\hat{b}_{p,q}$ ($\hat{b}_{p,q}^{\dagger}$) is the annihilation (creation)
operator at the lattice site $(p,q)$, $\hat{n}_{p,q}$ is the number operator,
and $U$ is the on-site interaction strength. The chemical potential $\mu$ 
fixes the total number of particles in the system and is suitable for 
studies with a mean-field in the grand-canonical ensemble. The ground state 
of the system can be obtained using the mean-field methods which decouples 
the bi-linear operators of neighboring lattice sites in the hopping term.


\subsection{Mean-field methods}
In the SGMF method, the annihilation and creation operators are decomposed 
in terms of a mean field and a fluctuation operator as 
$\hat{b}_{p,q} = \phi_{p, q} + \delta \hat{b}_{p, q}$ 
\cite{rokhsar_91, sheshadri_93}. Here 
$\phi_{p, q} = \langle \hat{b}_{p,q} \rangle$ is the mean-field and 
given by the expectation value of the annihilation operator with respect to 
the ground state of the system. This approximation allows the mean-field 
Hamiltonian of the system to be written as a direct sum of 
the mean-field Hamiltonians of the single sites 
\begin{eqnarray}
   \hat{h}_{p,q} =  &&-\Big( J
                    \phi^{*}_{p+1, q}\hat{b}_{p, q} 
                    + J \phi^{*}_{p, q+1}\hat{b}_{p, q} + {\rm H.c.}\Big)
                    \nonumber\\ 
                    &&+ \frac{U}{2} \hat{n}_{p, q} (\hat{n}_{p, q}-1) - 
                    \mu\hat{n}_{p, q}.
  \label{hamss}  
\end{eqnarray}
The ground state wavefunction can be expressed a product of the 
wavefunctions at each of the lattice sites by employing the Gutzwiller ansatz 
\begin{eqnarray}
   \ket{\Psi}_{GW} = \prod_{p,q} \ket{\psi}_{p,q}
                   = \prod_{p,q} \sum_{n=0}^{N_b} c_n^{p,q} \ket{n}_{p,q}.
  \label{gs_ss}
\end{eqnarray}
Here, $\ket{n}_{p,q}$ forms the Fock-space basis with $N_b$ representing the 
maximum allowed occupation number, and $c_n^{p,q}$s are the coefficients of 
the basis state $\ket{n}_{p,q}$.

To better account the correlation effects in the system, the CGMF method is 
used. In this method the lattice is tiled with clusters and for the
hopping within the same cluster or intra-cluster hopping, the hopping term 
of the BHM Hamiltonian is calculated exactly. However, inter-cluster hopping
is calculated using the mean-field. Like in the case of SGMF, the Hamiltonian 
of the system can be written as a direct sum of the  cluster Hamiltonians
and the mean field terms which couple the clusters \cite{luhmann_13}.
The ground state Gutzwiller wavefunction can then be defined as the direct
product of cluster wavefunctions expressed in the coupled basis. For the 
cluster of size $M\times N$, the cluster wavefunction is 
\begin{equation}
   \ket{\psi_c} = \sum\nolimits_{n_1,n_2...,n_{MN}}
                  C_{n_1,n_2..,n_{MN}}\ket{n_1,n_2...,n_{MN}},
\end{equation}
where $n_1,n_2..,n_{MN}$ are the occupancies at the sites within the cluster 
which are labelled with indices $1, 2, ... MN$. The equilibrium ground state 
is obtained numerically in a self-consistent iterative process. Starting with 
a random initial guess state, the Hamiltonian matrix is constructed 
and diagonalized to obtain the ground state wavefunction. From this 
wavefunction the mean-field is calculated which is then used as an improved 
guess. The iterations are performed till the mean-field 
converges \cite{bandyopadhyay_19, bai_18}.

The dynamics of the initial equilibrium state is studied by temporal 
evolution of the wavefunction according to the time-dependent Gutzwiller 
equations. These equations are essentially the Euler-Lagrange equations of 
motion derived from the Lagrangian of the system. The time-dependent 
Gutzwiller equations form a set of coupled partial-differential equations 
in terms of the dynamical coefficients of the wavefunction. These set of 
equations are solved using the fourth order Runge-Kutta (RK4) method.

\subsection{KZ Scaling for MI-SF QPT}

For a quantum quench across the MI-SF QPT, we choose an initial state in the
MI regime and quench the tunneling strength $J$ in time till it enters the
SF phase domain. The quench protocol is chosen as linear and is given by
\begin{equation}
   J(t) = J_i + (J_c -J_i) \frac{t + \tau_Q}{\tau_Q}.
\end{equation}
Here $J_i$ and  $J_c$ are the values of the quench parameter at the beginning
of the quench and at the critical point of the MI-SF phase boundary.
The different quench rates are realised by different values of the 
constant $\tau_Q$. Here, we have set $U =1$ and all system parameters are 
measured in this unit. The initial MI state evolves in response to the quench 
in parameter $J(t)$ across the criticality and enters into the Superfluid 
phase regime. The time instant at which the dynamically evolving state 
becomes a superfluid state is marked as time $\hat{t}$ and this happens 
after $t=0$ corresponding to the criticality $J_c$. This can be understood 
from the KZM according to which the critical slowing down near the continuous 
phase transition results in non-adiabatic evolution under the quench in the 
neighborhood of the critical point. In the KZM, the evolution of the state is 
adiabatic far from the criticality and impulse around the critical point. The 
adiabatic to impulse crossover time $\hat{t}$ is assumed to be the instant 
when the system relaxation time equals the inverse transition 
rate \cite{delCampo_14}. The power-law divergence of the equilibrium 
correlation length and relaxation time gives the KZM scaling laws for the 
defect density at the crossover time in terms of the critical exponents. The 
crossover time $\hat{t}$ scales as a power law with the quench 
rate $\tau_Q$ and is given by
\begin{equation}
   \hat{t} \propto \tau_{Q}^{\frac{\nu z}{1 + \nu z}}.
  \label{that_scale}
\end{equation}
Here, $\nu$ and $z$ are the equilibrium critical exponents. The defect density
at time $\hat{t}$ also scales as a power law
\begin{equation}
   \hat{N}_{\rm v}  \propto \tau_{Q}^{\frac{-(D-{\rm def}) \nu}{1 + \nu z}},
  \label{def_scale}
\end{equation}
where, $D$ is the system dimension and ${\rm def}$ is the dimension of the 
defects.

The superfluid state is characterised by the non-zero value of the average SF
order parameter 
\begin{equation}
   \phi = \frac{1}{N_s}\sum_{p,q} |\phi_{p,q}|,
\end{equation}
where, $N_s$ is the total number of lattice sites. Around the crossover time,
the system enters in SF phase and $\phi$ starts growing. The spontaneous
symmetry breaking at the phase transition, related to the global U(1) symmetry
breaking, leads to the production of topological defects in the system. This
arises because of the local choices of the order parameter when the 
symmetry breaking occurs at the phase transition. This results in the 
quenched state in the SF phase having numerous topological defects. For the 
case of point defects, vortex or antivortex, the conservation of angular
momentum implies the creation of vortex and anti-vortex occurs in pairs. By
definition, the vorticity around a vortex and anti-vortex are $+1$ and $-1$,
respectively, the total defects in the system can be counted as absolute 
sum of the vorticity over all the lattice site
\begin{eqnarray}
   N_{\rm v}    &=& \sum_{p,q} |\Omega_{p,q}|, \nonumber \\
   \Omega_{p,q} &=& \frac{1}{4}\big [\sin(\theta_{p+1,q} - \theta_{p,q})
                   + \sin(\theta_{p+1,q+1} - \theta_{p+1,q})
                                         \nonumber \\
                && -\sin(\theta_{p+1,q+1} - \theta_{p,q+1}) -
                    \sin(\theta_{p,q+1} - \theta_{p,q})\big],
                                         \nonumber \\
\end{eqnarray}
where $\theta_{p,q}$ is the phase of the SF order parameter $\phi_{p,q}$.


\section{Results and discussion} \label{sec_results}

We study the quantum quench across the MI-SF phase transition across the
MI(1) lobe. As mentioned earlier, we perform a linear quench of the hopping 
strength $J$ starting from the MI(1) state at $J_i = 0$ as the initial state 
and quench it across the MI-SF phase boundary into the SF regime. We study 
the dependence of the cross-over time $\hat{t}$ and defect density at 
$\hat{t}$ on the quench rate. The initial state corresponds to a equilibrium 
state at $J_i$. To this equilibrium state, we add small random fluctuations 
in the wavefunction. This is done to mimic the quantum fluctuations which 
drives the quantum phase transition. We add random uni-variate phase 
fluctuation ($0$ to $2\pi$) in the coefficient of the dominating basis 
states. Afterwards, we also introduce a random uni-variate density 
fluctuation of the order $10^{-3}$ in the coefficients of all the basis 
states. This completes the preparation of the initial state, and it is 
then evolved in time. The dynamical state is computed by solving the 
time-dependent Gutzwiller equations using the RK4 method. The quench is 
terminated in the superfluid regime. In the RK4 method we choose the time 
step $\Delta t = 0.005$, and note that the evolution is the same compared 
with the lower values of $\Delta t = 0.001$. To obtain good statistics of 
the defects produced during the quench, we choose a large system size of
$96 \times 96$ and employ the periodic boundary conditions along both the 
spatial dimensions. The dynamical properties of the state are calculated by 
taking a sample average of 40 different realizations. We study the quench 
at constant $\mu=0.3$ below the tip of the Mott lobe and for $\mu=0.4$ which 
corresponds to the tip of the MI(1) lobe. The studies are performed using 
the SGMF and CGMF methods and the results are discussed below.

\subsection{Quench across QPT below the MI(1) lobe ($\mu = 0.3$)}

The initial state is chosen as the MI(1) state on a $96 \times 96$ square
lattice and is obtained as an equilibrium solution for the system parameters
$\mu = 0.3$ and $J = 0$. This initial state is dressed with the fluctuations 
as mentioned earlier and is then time evolved according to the linear quench
protocol. Below we first discuss the properties of the quenched state 
using the SGMF method, and latter, compare and contrast the findings with 
the CGMF method.


\subsubsection{SGMF}

\begin{figure}[t]
   \includegraphics[width=6cm]{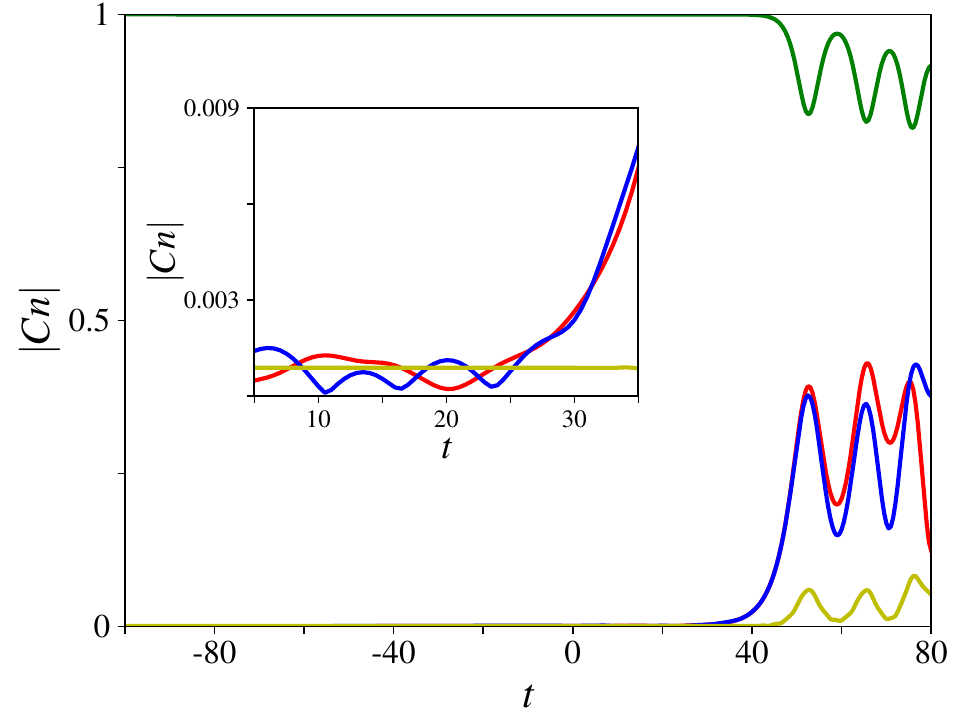}
   \caption{Dynamical evolution of the absolute magnitude of the
            coefficients of the basis states with SGMF for $\tau_Q = 100$.
	    Green color curve corresponds to unit filling basis sate,
	    red color curve corresponds to the vacuum state, while
	    blue and yellow color curve corresponds to the basis states 
	    with two and three particles respectively.}
  \label{sgmf_cn}
\end{figure}

\begin{figure}[t]
   \includegraphics[width=6cm]{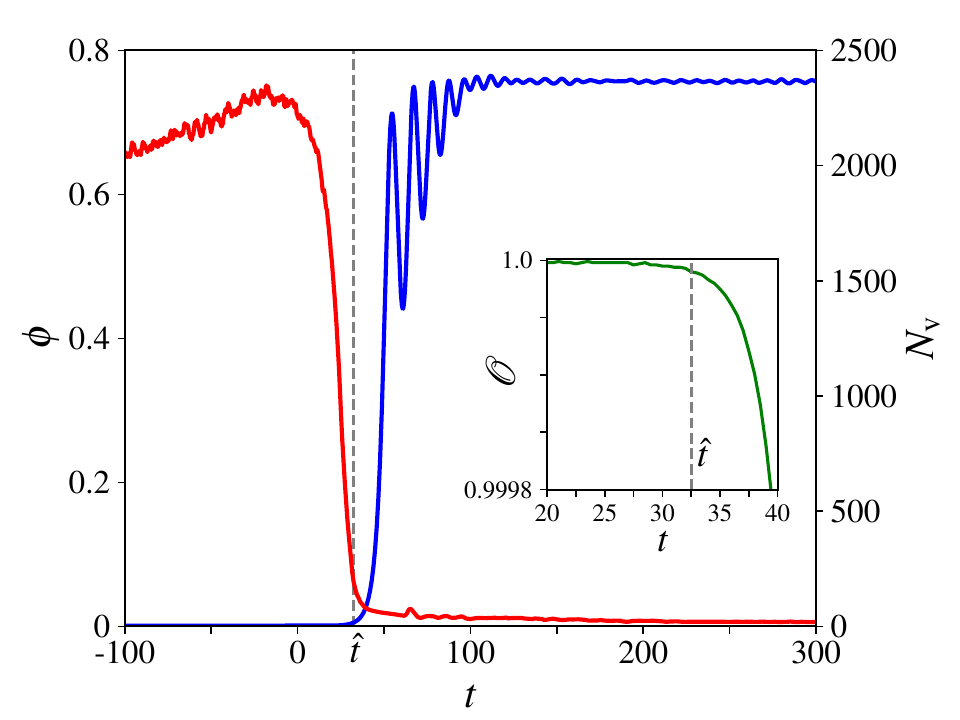}
   \caption{SF order parameter $\phi$ (blue) and defect density (red)
            across the phase transition for $\tau_Q = 100$. After $t=100$
	    quench is stopped and the state evolves freely.
	    The crossover time $\hat{t}$ is marked with dashed grey line.
	    The overlap measure ${\mathcal O}$ used to identify the
	    crossover time is shown in the inset.}
  \label{temp_evol}
\end{figure}

The initial MI(1) state at equilibrium is obtained using the SGMF method
with a sufficiently large cutoff for the single-site occupancy $N_b = 6$. 
The Fig.{\ref{sgmf_cn}} shows the temporal evolution of the magnitude of the 
coefficients of various Fock-states. The green color curve corresponds to the
coefficient of $\Ket{n=1}_{p,q}$ Fock state and the red, blue and yellow 
color curves corresponds to the Fock state with 0, 2 and 3 occupancies 
respectively. Starting from $t= -\tau_Q$, the adiabatic evolution continues 
and the quenched state remains in the MI(1) state as can be seen from 
green curve with value 1. However, the quenched state remains MI(1) state 
for some time even after crossing the criticality at $t = 0$. This is
in agreement with the assumption of impulse regime where the state 
is frozen. The corresponding fluctuation in the coefficients shown in the 
plot are of the order of $10^{-3}$, which is due to the initial density
fluctuations in the dressed state. The quenched state, which is frozen in 
the impulse domain, starts to evolves dynamically only after the quench 
parameter $J$ is sufficiently away from $J_c$. The time evolution of the SF 
order parameter of the quenched state is shown in Fig.{\ref{temp_evol}} for 
$\tau_Q = 100$. As shown in the figure with blue curve, $\phi$ is zero 
initially as it should be in the MI state. However it remains zero even after 
passing the criticality at $t = 0$ with $J(t=0) = J_c$, and shows growth 
only after a certain time instant $\hat{t}$. After $\hat{t}$, the 
SF order parameter grows rapidly and followed by an oscillating trend
with decaying amplitude. Finally, after $t=\tau_Q$ when the quench 
terminates, the system is allowed to evolve freely. In this domain
the value of $\phi$ quickly saturates. The temporal evolution
of the defect density ($N_{\rm v}$) is shown in Fig.{\ref{temp_evol}} with the
red curve. At initial times, the system has a large number of vortices and
anti-vortices owing to the introduction of phase fluctuations in
the initial dressed state. However, $N_{\rm v}$  decays rapidly after the
criticality ($t = 0$) due to the initiation of phase coherence in the system.
The KZM scaling laws predicts the power-law scaling of the defect density at
$\hat{t}$ with the quench rate. To identify $\hat{t}$, we use the overlap
of the wavefunction at any time with the wavefunction at time $t = 0$,
\begin{equation}
   {\mathcal O}(t) = |\langle\psi(0)|\psi(t)\rangle|.
\end{equation}
This definition is motivated from the fact that the dynamical evolution of the
state is frozen in the impulse domain as observed in Fig.{\ref{sgmf_cn}}.
The KZM predicts the overlap to be unity till $\hat{t}$, and afterwards it 
can deviate from unity. This is because, in the ``impulse" regime the dynamical
wavefunction can change with time only upto a phase factor. The deviation of 
overlap from unity signals the crossing of $\hat{t}$ and occurs when the value 
of overlap is $\approx 0.99999$. This value depends on the magnitude of the 
density fluctuations introduced in the dressed initial state. Consequently, 
we choose a threshold value of $0.99999$ for the overlap as a locator of 
$\hat{t}$. The overlap measure is shown in the inset of the 
Fig.{\ref{temp_evol}}.

\begin{figure}[t]
   \includegraphics[width=6cm]{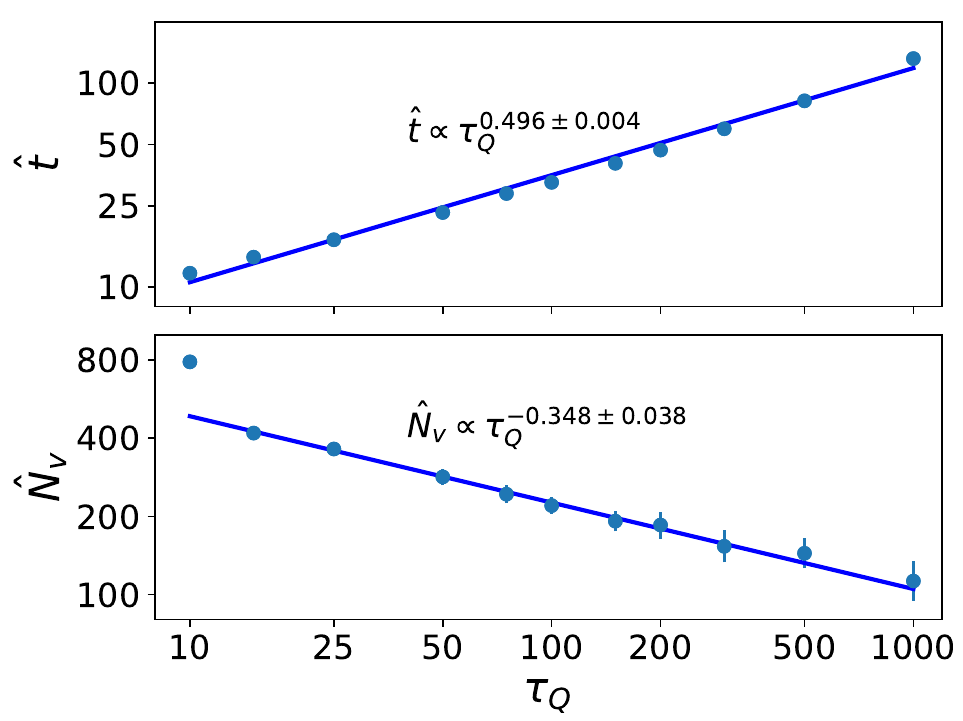}
   \caption{Power law scaling of $\hat{t}$ and $\hat{N}_{\rm v}$ with $\tau_Q$.
	    The crossover time is located using the overlap criterion.}
  \label{kz_scaling1b1}
\end{figure}

\begin{figure}[t]
   \includegraphics[width=6cm]{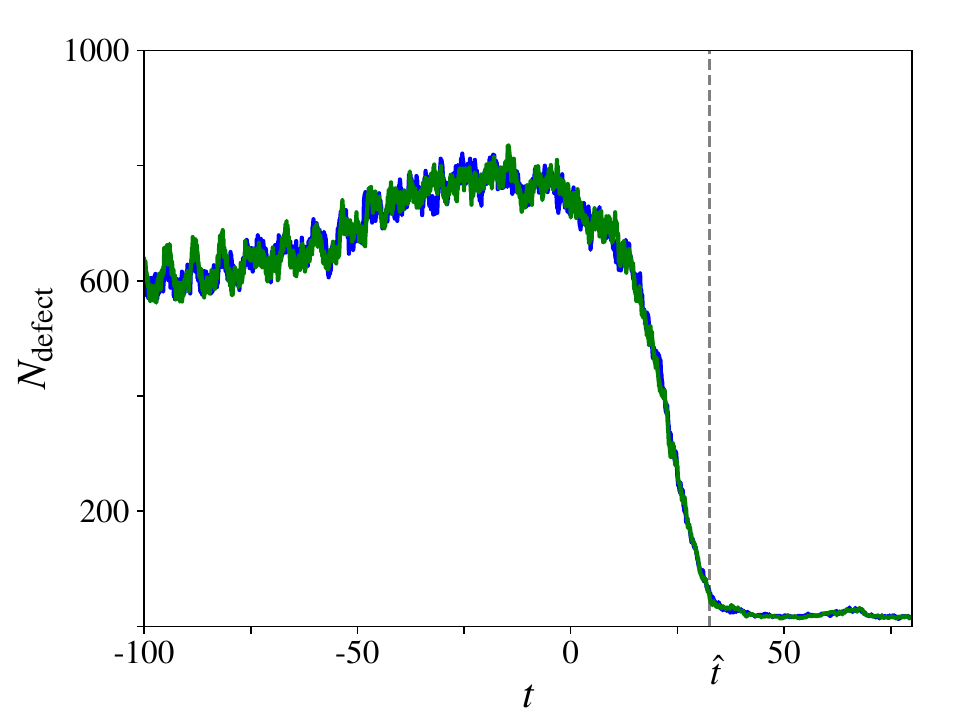}
   \caption{Plot of topological defects in the quenched state as a function
            of time. Blue curve shows the number of vortices with a
            vorticity $\Omega_{p,q} > 0.5$, while the green curve shows
            the  number of anti-vortices with a vorticity 
            $\Omega_{p,q} < -0.5$. The blue and green curves sit on top
            of each other, with almost equal number of vortex and anti-vortex
            in the quenched state.}
  \label{vort_antivort}
\end{figure}

For different values of the quench rate $\tau_Q$, we note the crossover time
$\hat{t}$ and the corresponding defect density $\hat{N}_v$, and the results
are shown in Fig.{\ref{kz_scaling1b1}}. It can be seen that a power-law scaling
exists, but the power-law behaviour shows deviations at the extreme values of
quench rates and is much pronounced in the plot of defect density for very fast
quenches. However, for very slow quenches, the deviations in the defect density
across the samples gives large error bars in the plot. So the power-law
exponents are determined by fitting the data points with a power-law in the
regime $\tau_Q \in [25, 300]$. From Fig.{\ref{kz_scaling1b1}}, we note that the
crossover time $\hat{t} \propto \tau_Q^{0.496 \pm 0.004}$ and the defect density
$\hat{N}_{\rm v} \propto \tau_Q^{-0.348 \pm 0.038}$. A comparison of these
power law exponents with the Eq.\ref{that_scale} and Eq.\ref{def_scale} gives
$\nu z/(1+ \nu z) = 0.496 \pm 0.004$ and 
$(D-{\rm def}) \nu/(1 + \nu z) = 0.348 \pm 0.038$. A important point to note 
here is that the topological defects are produced as vortex anti-vortex pairs 
and constitute as a 1-D defect, that is ${\rm def} = 1$ \cite{zhou_20}.
This can be seen from Fig.{\ref{vort_antivort}}, where at 
$t = -\tau_Q$ the initial dressed state have equal number of vortices and 
anti-vortices owing to the random phase fluctuations. As the quench 
progresses, the defects are created or annihilated in pairs, thereby 
maintaining the equality between their numbers.  Thus, we obtain the 
critical exponents $\nu = 0.69 \pm 0.11$ and $z = 1.42 \pm 0.17$.

\begin{figure}[t]
   \includegraphics[width=6.0cm]{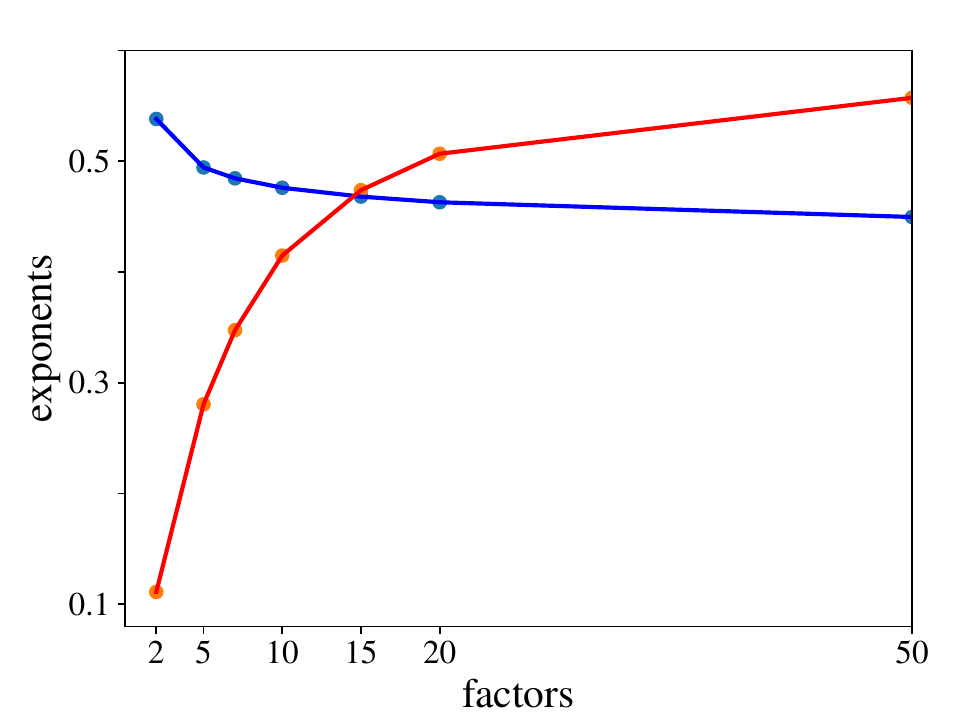}
   \caption{Scaling exponents for $\hat{t}$ (blue) and $\hat{N}_{\rm v}$
            (red) for various choices of the growth factor 
	    $\phi(\hat{t}) / \phi(0)$.}
  \label{grow_fac1b1}
\end{figure}

It should be noted that there are various works in which the crossover time
$\hat{t}$ is chosen as the time instant where average SF order parameter
becomes twice of its value at criticality $\phi(\hat{t}) = 2\phi(0)$
and around this time $\phi$ starts to increase rapidly 
\cite{shimizu_misf_18, zhou_20}.
However, this choice of the growth factor $2$ is arbitrary. So, we have also 
checked the power law scaling for various choices of the growth factor 
$\phi(\hat{t}) / \phi(0)$ and investigated the form of the power law. We
find that the variation in the power-law exponents is large and depends on 
the chosen value of the growth factor. This is reported in 
Fig.{\ref{grow_fac1b1}} where it can be seen that the variation in the 
power-law exponents is absent only for large values of the growth 
factor $\sim 20$. This is because around this time, the rate of
increase in order parameter is quite large and different choice of the growth 
factors corresponds to very close values of time $t$. Comparing this with the
overlap criterion discussed earlier, we find that the growth factor of $7$ 
gives reasonably close values for the power-law exponents obtained with the 
overlap protocol in our studies. Next, we study the dynamics with the CGMF 
method as described below.


\subsubsection{CGMF}

\begin{figure}[t]
   \includegraphics[width=6cm]{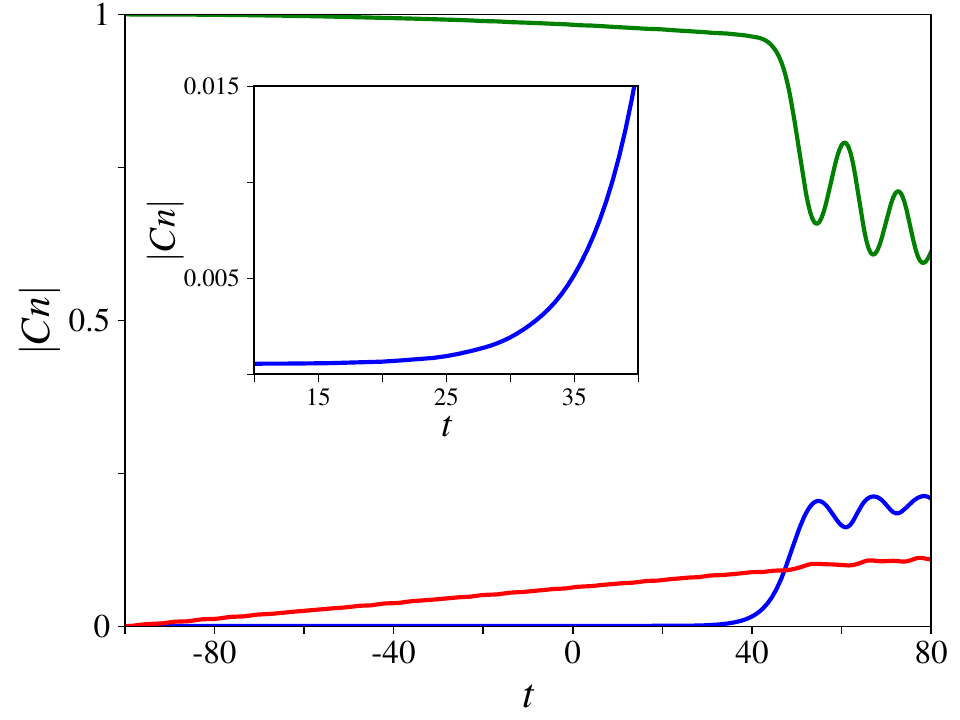}
   \caption{Dynamical evolution of the absolute magnitude of the
            coefficients of the basis states with $2 \times 2$ cluster
	    CGMF for $\tau_Q = 100$.
	    Green color curve corresponds to uniform unit filling basis sate,
	    blue color curve corresponds to the basis state with an
	    extra particle/hole on top of uniform unit filling, and
	    red color curve corresponds to 4-particle states different
	    from uniform unit filling.}
  \label{evol_cn}
\end{figure}

\begin{figure}[t]
   \includegraphics[width=6cm]{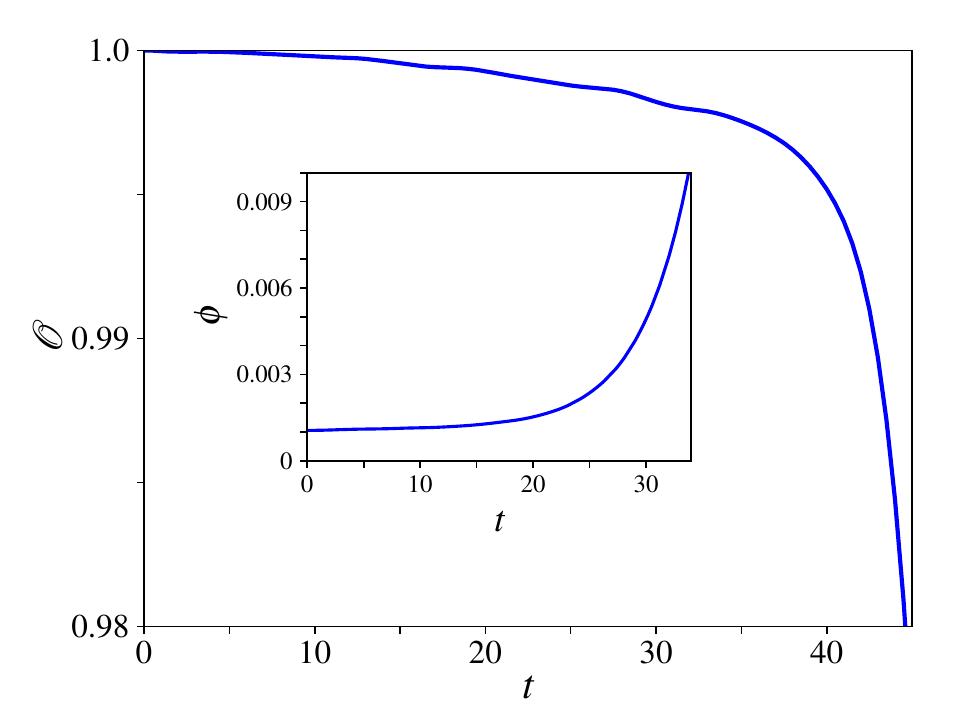}
   \caption{Overlap measure with $2 \times 2$ cluster CGMF for 
            $\tau_Q = 100$. Overlap doesn't show a sharp decay and is
	    not-suitable for locating $\hat{t}$. The average SF order
	    parameter is shown in the inset.}
  \label{olap_cgmf}
\end{figure}

With the CGMF method, we have chosen $2\times 2$ (and $2\times 3$) clusters
to tile the $96 \times 96$ lattice. Allowing the single-site occupancy to be 
atmost 2, the coupled basis states for the $2\times 2$ cluster now comprise
of total $81$ possible configurations ($729$ basis states for $2\times 3$).
It is difficult to use large clusters in studying the dynamics as the
associated size of the Fock space grow exponentially large with increasing 
cluster sizes. To circumvent this problem by reduction of basis states 
appropriately is not a good idea, as in dynamics the excitations are 
introduced in the state near the phase transition (the extra energy being 
proportional to the fastness of the quench rate). We have checked this 
behaviour in our studies. We also observe the intra-cluster dynamics to be 
present near the critical point which is absent in the SGMF method. This 
intra-cluster dynamics can be seen in Fig.{\ref{evol_cn}} where the absolute 
magnitude of the coefficients for some of the basis states are plotted for 
studies with $2\times 2$ cluster. The red color curve corresponding to the 
coefficient of basis states with 4 particles but different from the uniformly 
filled basis state keeps growing from the beginning of quench. And the 
strength of the uniform unit filling basis state (green color curve) 
decreases with time. Owing to the better inclusion of correlation effects in 
the wavefunction due to the intra-cluster dynamics, the CGMF method captures 
the evolution of the dynamical wavefunction in the ``impulse" regime of the 
KZM more accurately. Naturally, in the overlap protocol to determine 
$\hat{t}$ introduced earlier, the deviation from unity would arise earlier 
because of the intra-cluster dynamics. This can  be seen without ambiguity
from the overlap as shown in Fig.{\ref{olap_cgmf}}. Thus, the overlap protocol 
is not suitable for locating the crossover time $\hat{t}$.
The time $\hat{t}$ corresponds to the time at which the SF order parameter
develops and inter-cluster dynamics starts. This time can be read from the
Fig.{\ref{evol_cn}} as the time when the coefficients of the basis states
corresponding to an extra particle/hole on top of uniform unit filling 
(blue color curve) starts growing. This happens at $t \sim 30$ and the SF order
parameter around this time grows rapidly compared to its value at criticality
as can be seen from the inset of Fig.{\ref{olap_cgmf}}.

\begin{figure}[t]
   \includegraphics[width=6cm]{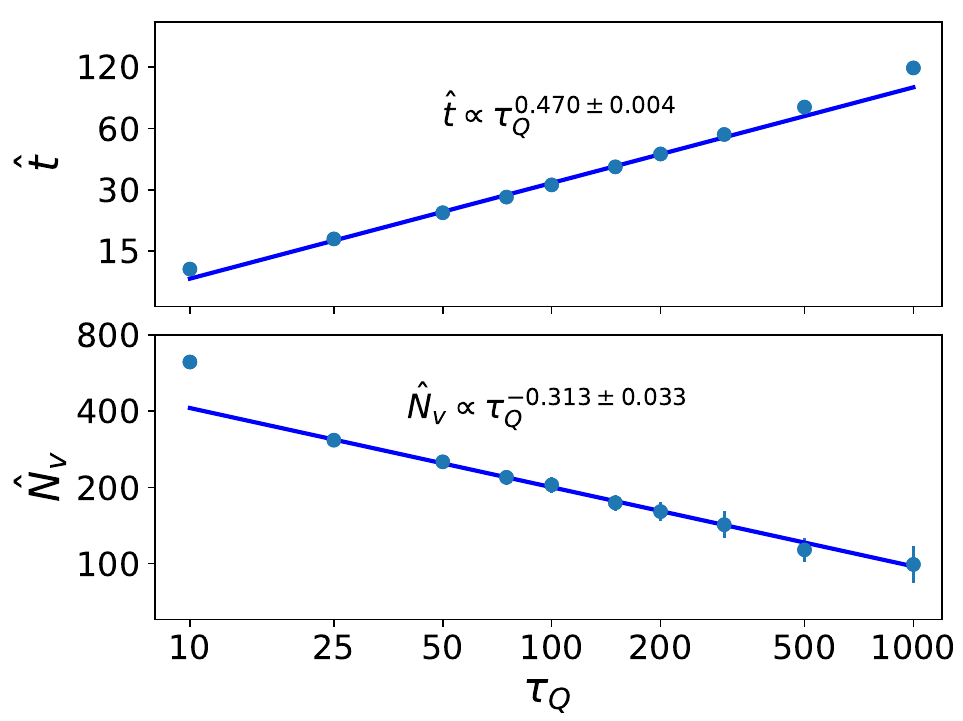}
   \caption{Power law scaling of $\hat{t}$ and $\hat{N}_v$ with CGMF
            using 7-fold growth factor of $\phi$ as indicator for
	    $\hat{t}$ for cluster size of $2\times 2$.}
  \label{2b2clus_scale}
\end{figure}

\begin{figure}[t]
   \includegraphics[width=6cm]{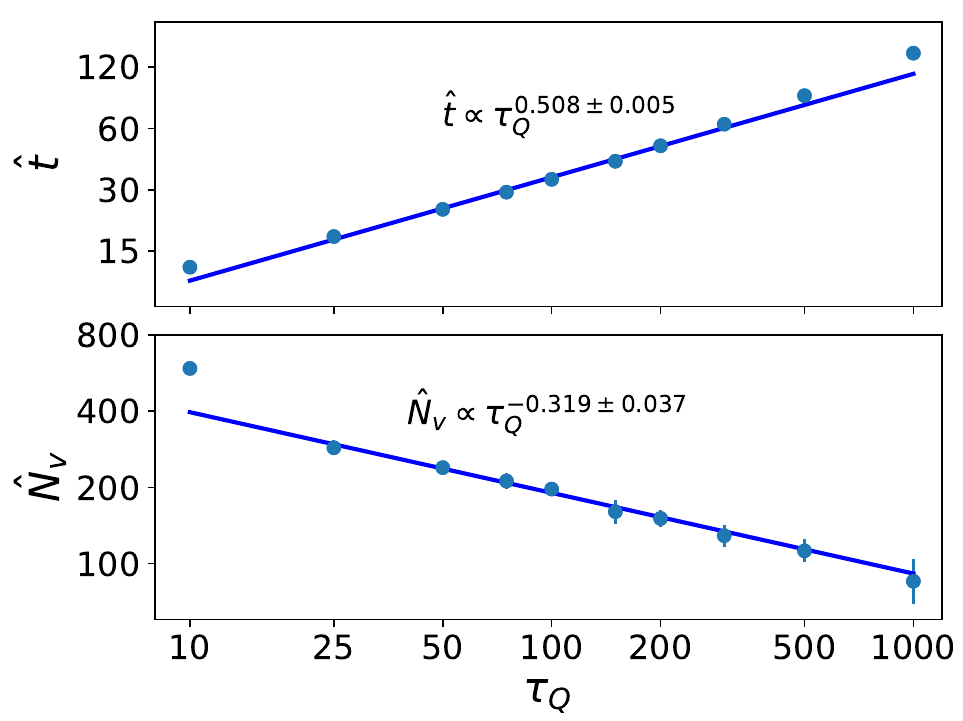}
   \caption{Power law scaling of $\hat{t}$ and $\hat{N}_v$ with CGMF
            using 7-fold growth factor of $\phi$ as indicator for
	    $\hat{t}$ for cluster sizes $2\times 3$.}
  \label{2b3clus_scale}
\end{figure}

\begin{figure}[t]
   \includegraphics[width=4.2cm]{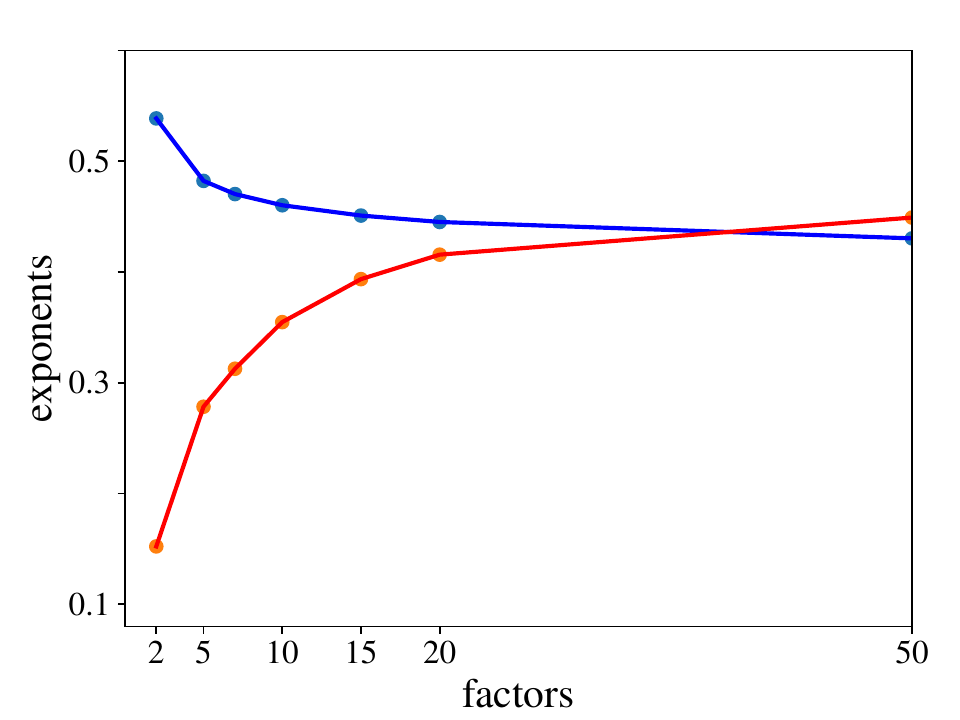}
   \includegraphics[width=4.2cm]{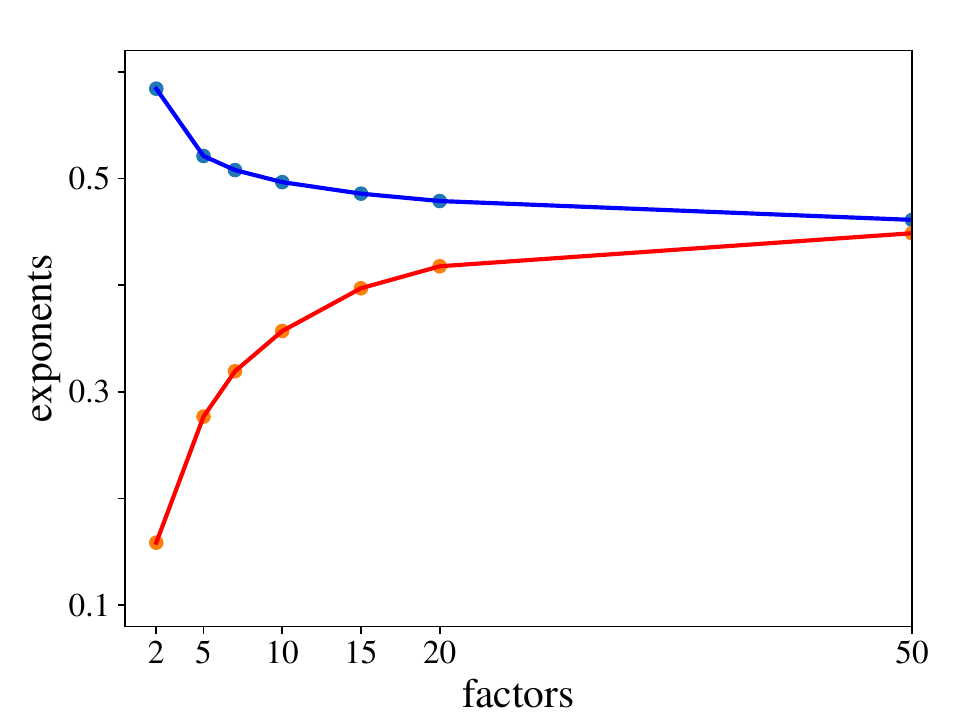}
   \caption{Scaling exponents for $\hat{t}$ (blue) and
            $\hat{N}_v$ (red) for various choices of the growth factor
	    $\phi(\hat{t}) / \phi(0)$ with CGMF of cluster $2 \times 2$ (left)
	    and $2 \times 3$ (right).}
  \label{clus_factors}
\end{figure}

Motivated by the SGMF studies, we use the definition of 7-fold growth in
$\phi$ as an identifier of $\hat{t}$ to investigate the KZM scaling laws with 
the CGMF method. The crossover time and defect density show power law scaling 
with the quench rate as shown in Fig.{\ref{2b2clus_scale}} and 
Fig.{\ref{2b3clus_scale}} for CGMF with $2 \times 2$ and $2 \times 3$ 
clusters respectively. As can be seen, the power law behaviour deviates for 
very fast and very slow quenches. The power law scaling is thus evident in 
the regime $\tau_Q \in [25, 300]$. It is to be noted that a different choice 
of the growth factor $\phi(\hat{t})/\phi(0)$ for locating crossover time 
leads to different scaling exponents similar to the SGMF case as shown 
in Fig.{\ref{clus_factors}}. Using the KZ scaling laws 
$\hat{t} \propto (\tau_Q)^{\frac{\nu z}{1+ \nu z}}$, and
$\hat{N}_{\rm v} \propto (\tau_Q)^{\frac{\nu}{1+ \nu z}}$, we obtain the
critical exponents $\nu$ and $z$. A comparison of the critical exponents 
obtained with the SGMF and CGMF methods for 7-fold growth in $\phi$ as a 
locator for $\hat{t}$ is shown in Table{\ref{that_expo}} where $1\times 1$ 
cluster means the SGMF method. The critical exponent $\nu$, which is 
associated with the divergence of the correlation length $(\xi)$, shows a 
marginal change with the SGMF and CGMF method but doesn't show any trend 
with the cluster sizes. However the critical exponent $z$, which is 
associated with the divergence of the relaxation time $(\tau)$, shows a 
increasing trend with the increase in the cluster size. These calculated 
exponents can be compared with the equilibrium mean-field exponents 
$\nu = 1/2$ and $z = 2$.

\begin{table}[t]
   \begin{tabular}{ |c|c|c|c|c|c|}
   \hline
   cluster        &\hspace{0.8cm} $1\times1$     \hspace{0.8cm}
   	       &\hspace{0.8cm} $2\times2$     \hspace{0.8cm} 
   	       &\hspace{0.8cm} $2\times3$     \hspace{0.8cm} \\
   \hline
   $\frac{\nu z}{1+\nu z}$ & $0.484 \pm 0.003$ & $0.470 \pm 0.004$ 
                                               & $0.508 \pm 0.005$ \\
   \hline
   $\frac{\nu}{1+\nu z}$   & $0.347 \pm 0.034$ & $0.313 \pm 0.033$ 
                                               & $0.319 \pm 0.037$ \\
   \hline
   $z$                     & $1.39 \pm 0.15$   &  $1.50 \pm 0.17$ 
                                               & $1.59 \pm 0.20$ \\
   \hline
   $\nu$ 	                & $0.67 \pm 0.09$   &  $0.59 \pm 0.09$
                                               & $0.65 \pm 0.11$ \\
   \hline
   \end{tabular}
   \caption{Critical exponent for $\hat{t}$ and $\hat{N}_v$ for various
            cluster sizes, for QPT at $\mu = 0.3$.}
   \label{that_expo}
\end{table}


\subsection{Quench across the tip of MI(1) lobe}

\begin{figure}[t]
   \includegraphics[width=6.0cm]{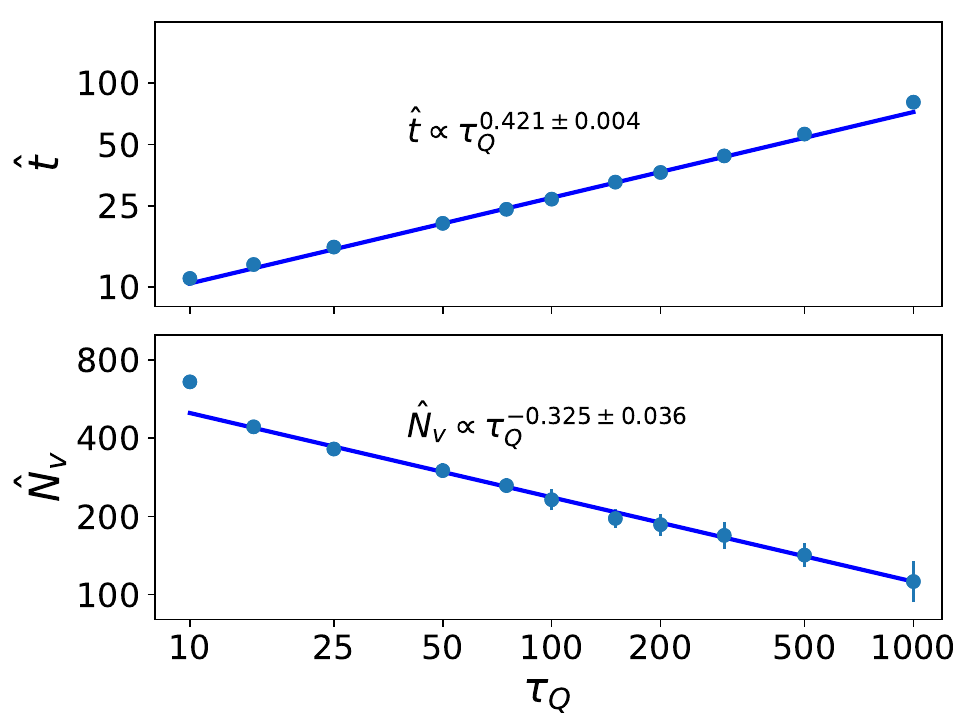}
   \caption{Power law scaling of $\hat{t}$ and $\hat{N}_v$, calculated with
            SGMF using the overlap criterion for locating $\hat{t}$.}
  \label{tip_kz_scale1b1}
\end{figure}

We quench the initial MI state across the multi-critical point at the tip of
the MI(1) lobe ($\mu= 0.4$). This continuous QPT corresponds to the 3D XY 
model universality class. As described earlier, the initial state corresponds 
to the equilibrium MI(1) state and is dressed with random fluctuations. The 
dressed state is then time evolved by quenching $J$ across the QPT into the 
SF regime. Similar to the previously discussed case of quantum quench below 
the tip of the Mott lobe ($\mu= 0.3$), we observe a delayed growth of the 
average SF order parameter in the quenched state. And the quenched state has 
topological defects (vortex-antivortex pairs). With the SGMF method, we find 
impulse regime near the criticality and thus utilize the overlap protocol 
for the identification of crossover time $\hat{t}$. We observe power law 
behaviour of the cross-over time 
$\hat{t} \propto \tau_Q^{0.421 \pm 0.004}$ and the defect density
$\hat{N}_{\rm v} \propto \tau_Q^{-0.325 \pm 0.036}$ and is shown in
Fig.{\ref{tip_kz_scale1b1}}. The power law fitting is done using data in the
regime $\tau_Q \in [25,300]$. We also observe that the overlap protocol for 
locating $\hat{t}$ is consistent with the 7-fold growth in $\phi$ compared to
the value at criticality. The power-law exponents are different from the
previously obtained exponents at $\mu = 0.3U$. This is expected since the 
QPT at the tip belongs to a different universality class with the critical 
exponents given by the 3D XY model. From the power law exponents of cross-over
time and the defect density, we find $\nu =0.56 \pm 0.09$ and 
$z =1.30 \pm 0.16$ which should be compared with the equilibrium values 
$\nu = 2/3$ and $z = 1$.

\begin{figure}[t]
   \includegraphics[width=6.0cm]{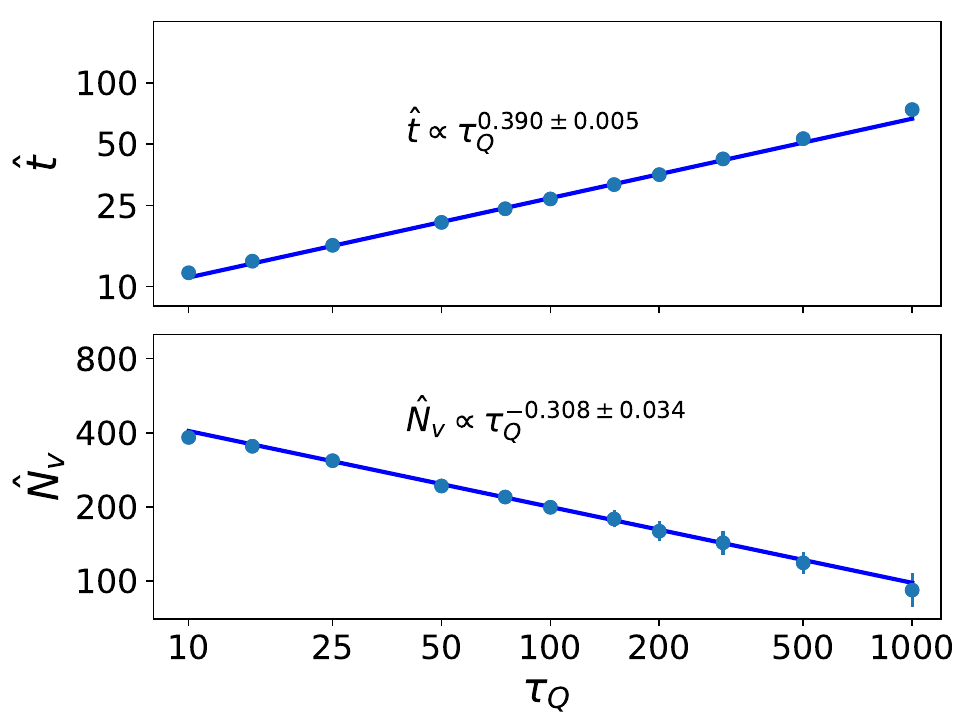}
   \caption{Power law scaling of $\hat{t}$ and $\hat{N}_v$, with $2\times 2$
	    cluster}
  \label{tip_kz_scale2b2}
\end{figure}

\begin{figure}[t]
   \includegraphics[width=6cm]{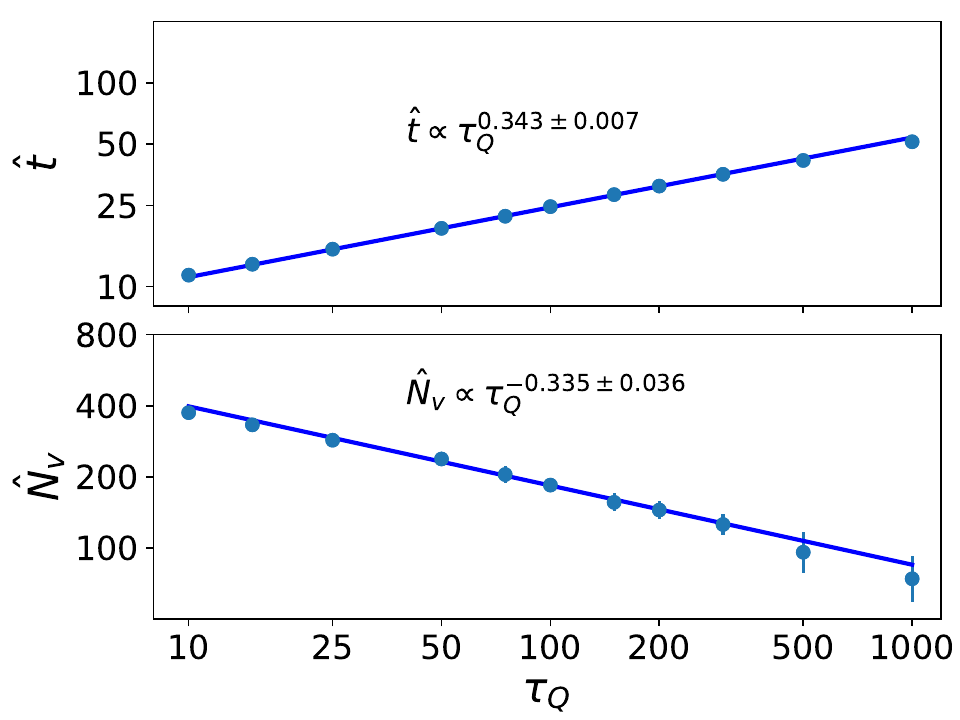}
   \caption{Power law scaling of $\hat{t}$ and $\hat{N}_v$, with $2\times 3$
            cluster}
  \label{tip_kz_scale2b3}
\end{figure}

With CGMF we use $2\times 2$ and $2\times 3$ clusters to study the evolution of
the quenched state across the QPT. As seen previously for QPT at $\mu = 0.3$,
we observe that the dynamical state evolves in the ``impulse regime".
Utilizing the 7-fold growth in $\phi$ as locator for crossover time, the power
law scalings for cross-over time and defect density are shown in 
Fig.{\ref{tip_kz_scale2b2}} and Fig.{\ref{tip_kz_scale2b3}} for CGMF studies 
with $2\times 2$ and $2\times 3$ cluster respectively.
A comparison of the critical exponents with SGMF and CGMF methods, obtained
using the 7-fold growth in $\phi$ for locating $\hat{t}$ is shown in
Table{\ref{tip_expo}}. The results suggests the critical exponents
$\nu \sim 1/2$ and $z \sim 1$ which are close to the equilibrium critical
exponents for 3D XY model $\nu = 2/3$ and $z = 1$.

\begin{table}[t]
   \begin{tabular}{ |c|c|c|c|c|c|}
   \hline
   cluster                 &\hspace{0.8cm} $1\times1$     \hspace{0.8cm}
                           &\hspace{0.8cm} $2\times2$     \hspace{0.8cm} 
                           &\hspace{0.8cm} $2\times3$     \hspace{0.8cm} \\
   \hline
   $\frac{\nu z}{1+\nu z}$ & $0.407 \pm 0.004$ & $0.390 \pm 0.005$ 
                                               & $0.343 \pm 0.007$ \\
   \hline
   $\frac{\nu}{1+\nu z}$   & $0.335 \pm 0.032$ & $0.308 \pm 0.034$ 
                                               & $0.335 \pm 0.036$ \\
   \hline
   $z$                     & $1.21 \pm 0.13$   &  $1.27 \pm 0.16 $ 	
                                               & $1.02 \pm 0.13$ \\
   \hline
   $\nu$                   & $0.57 \pm 0.08$   &  $0.50 \pm 0.09$
                                               & $0.51 \pm 0.10$ \\
   \hline
   \end{tabular}
   \caption{Critical exponent for $\hat{t}$ and $\hat{N}_v$ for various
            cluster sizes, for QPT at $\mu = 0.4$.}
   \label{tip_expo}
\end{table}

\section{Conclusions} \label{sec_summary}

We have studied the quantum quench dynamics of ultracold bosons in a 2D
square optical lattice. The quenching of hopping amplitude drives the QPT
across the MI-SF phase boundary of BHM. We study the dynamical properties of
the quenched state from the perspective of KZM and have obtained the
power-law scaling of the crossover time and defect density with the quench
rate using the SGMF and CGMF methods. The power-law behavior, however, deviates
at very slow and fast quenches. The CGMF studies capture the evolution of
the quenched state in the ``impulse" regime of the KZM. The critical exponents
obtained from the dynamics are close to the equilibrium values.
In particular, the critical exponent $z$, which is associated with the 
divergence of the relaxation time, improves towards the equilibrium
value with higher cluster sizes of CGMF. This is expected as the larger
cluster size allows a large number of basis states for redistribution of the
population in the initial state and allows better relaxation dynamics.

\section{Acknowledgements}

The results presented in this paper were computed on Vikram-100, the 100TFLOP
HPC cluster and Param Vikram-1000 HPC cluster at Physical Research laboratory, 
Ahmedabad, India. DA would like to acknowledge support from the Science and
Engineering Research Board, Department of Science and Technology, Government
of India through Project No. CRG/2022/007099 and support from the UGC through
the SAP (DRS-II) project F.530/18/DRS-II/2018(SAP-I), Department of Physics,
Manipur University. The authors are grateful to Dr. Sukla Pal,
Dr. Kuldeep Suthar, Dr. Rukmani Bai and Dr. Soumik Bandyopadhyay for the
fruitful discussions.


\bibliography{ref}{}

\end{document}